# Magnetic properties of BiFeO$_3$ micro-cubes synthesized by microwave agitation

B. Andrzejewski[1*], K. Chybczyńska[1], K. Pogorzelec-Glaser[1], B. Hilczer[1], B. Łęska[2], R. Pankiewicz[2], P. Cieluch[3]

[1] *Institute of Molecular Physics, Polish Academy of Science, Smoluchowskiego 17, PL-60179 Poznań, Poland*

[2] *Adam Mickiewicz University, Faculty of Chemistry, Grunwaldzka 6, PL-60780 Poznań Poland*

[3] *The Research Centre of Quarantine, Invasive and Genetically Modified Organisms, Institute of Plant Protection – National Research Institute, Węgorka 20, PL-60318 Poznań, Poland*

*\* Corresponding author: and@ifmpan.poznan.pl*

In this report we present results of field-cooled and zero field-cooled magnetization measurements and investigation of aging and memory effect in bismuth ferrite multiferroic micro-cubes obtained by means of simple microwave synthesis procedure. It is found that difference between field-cooled and zero field-cooled magnetizations appears at the temperature of freezing of ferromagnetic domain walls. The decay of the magnetic moment vs. time described by power-law relation and the absence of memory effect are caused by domain growth mechanism rather than by the spin-glass phase. The negligible value of remnant magnetic moment indicates that bismuth ferrite compound exhibits low concentration of ferromagnetic domains and can be close to ferromagnetic to spin-glass transition.

## 1. Introduction

Bismuth ferrite BiFeO$_3$ (BFO) compound belongs to an interesting class of magnetoelectric (ME) multiferroic materials that exhibit a coexistence of mutually coupled ferroelectric and magnetic ordering. These ME materials have recently attracted attention of researchers all over the world because of their perspective applications in nanophysics, sensors, future electronic and multi-state memory devices [1-5].

BFO multiferroic compound shows a rhomboedrally distorted perovskite cell with space group *R3c* at room temperature. Ferroelectricity in BFO arises below the Curie temperature $T_C$=1143 K and is due to the ordering of lone pairs of Bi$^{3+}$ ions located at the A sites of the perovskite unit cell [6]. The ferroelectric ordering causes irreversible polarisation vs. electric field *P(E)* with a value of remnant polarisation $P_r$ ≈35 μC·cm$^{-2}$ at room temperature [4], close to the estimations from first-principles calculations [7, 8]. Magnetic properties of BFO are associated with Fe$^{3+}$ spins occupying the B sites in the perovskite unit cell. From the point of view of magnetic ordering, BFO is an antiferromagnet (AFM) below the Néel temperature $T_N$=643 K. The magnetic order in BFO is subjected to other transitions manifested near the temperatures 140 K and 200 K. The transition at 140 K is interpreted in terms of spin-glass phase appearance [9, 10] whereas the anomaly about 200 K is ascribed to spin reorientations assisted by magnon softening [11].

AFM ordering in BFO is a rather complex G-type structure, in which each spin of iron is surrounded by six antiparallel coordinated spins, with superimposed long-range incommensurate cycloidal modulation [12]. The spin cycloid can propagate along three equivalent vectors $q_1$=[1,-1,0], $q_2$=[1,0,-1] and $q_3$=[0,-1,1] (in pseudocubic notation) with the period of λ=62 nm. In the cycloid, the Fe$^{3+}$ spins rotate in the plane determined by one of the three propagation vectors *q* and the vector of spontaneous electric polarization $P_s$ in the direction [111]. The magnetoelectric interaction similar in the form to that of Dzyaloshinski-Moriya type causes small canting of spins out of the rotation plane, which produces a weak ferromagnetic moment (FM): $m \sim P_s \times L$ where $L$ is the antiferromagnetic vector. Cycloidal modulation brings about an analogous space-modulation of $m$ vector, spin density wave (SDW) and the absence of spontaneous magnetization [13]. However, the homogeneous G-type ordering with no cycloidal modulation can be established in BFO by anisotropic strains or a strong magnetic field and weak macroscopic FM with a theoretical magnitude of 0.02÷0.1 $\mu_B$/Fe can appear. Therefore, in BFO multiferroic probably the FM ordered and spin-glass phases can simultaneously exist.

The systems exhibiting both ferromagnetic and spin-glass properties like, for example chromium thiospinels, have been reported by Vincent et al. [14]. Depending on chromium content the thiospinels could be ferromagnet, spin-glass or ferromagnet followed by re-entrant spin-glass phase at lower temperature. It has been demonstrated that many properties of disordered FM phases and spin-glasses like aging, rejuvenation, magnetization dependence on history and response to AC magnetic field are indeed similar.

The aim of this work is to study the possible mutual interactions of the ordered FM and spin-glass phase in BFO and to answer the question which phenomenon brings a dominant effect on the magnetic properties: spin-glass or FM domain growth and pinning of the domain walls.

2. **Experimental**

A simple microwave synthesis procedure was used to obtain bismuth ferrite multiferroic micro-cubes. The micro-cubes were formed by means of microwave assisted hydrothermal Pechini process. The substrates for the synthesis, $Bi(NO_3)_3 \cdot 5H_2O$, $Fe(NO_3)_3 \cdot 9H_2O$, $Na_2CO_3$ were mixed and added to a KOH water solution (11 M). All chemicals were of analytical grade and were used without further purification. Then the mixture was poured into a few teflon reactors and processed in CEM Mars-5 microwave oven. The microwave reaction was carried out at 200 °C for 30 min. The as-obtained brown in coloration, fine powders were collected by decanting, rinsed by water and dried at 70 °C in an conventional oven. Some powders were also air calcined at 500 °C for 1h, however this treatment had no distinct effect on their crystallographic structure and physical properties.

The samples were characterized by powder X-ray diffraction (XRD) performed on an ISO DEBEYE FLEX 3000 instrument, using a Co lamp generating 0.17928 nm, wavelength. The morphology of the samples was examined by S3000N Hitachi Scanning Electron Microscope (SEM).

The magnetic measurements were performed by means of the Physical Property Measurement System (PPMS Quantum Design Ltd.). The VSM (Vibrating Sample Magnetometer) probe was fitted to PPMS and used for DC magnetic moment measurements. The powder samples were sealed in gelatine capsules.

3. **Results and Discussion**

Fig. 1 represents the X-ray diffraction pattern of as-obtained microwave-synthesized product. The vertical lines terminated by squares correspond to the simulated spectrum of $BiFeO_3$ (BFO) for X-ray radiation with the wavelength 0.17928 nm as used in the experiment. Comparing these two plots one can clearly see that the as-obtained microwave-synthesized product is in good crystalline BFO rhombohedral phase with the *R3c* space group and contains very low amount of $Bi_{25}FeO_{40}$ parasitic phase.

SEM microscopy revealed that the powders were composed of larger agglomerates formed by BFO almost regular micro-cubes (see Fig. 2). The slightly rounded corners of the cubes were due to the effect of high KOH concentration in solution, actually equal to 11 M. The mean size of the BFO micro-cubes was about 1 μm.

The zero-field cooling (ZFC) and field cooling (FC) temperature dependences of magnetization $M(T)$, are shown in Fig. 3. For the sake of simplicity, this figure presents ZFC and FC data for three values of the applied magnetic fields only. The ZFC measurements were performed on warming the sample, after prior cooling it to a low temperature in the absence of a magnetic field, followed by applying of the field of a given strength. The FC data were also recorded on warming, after previous cooling of the sample in a magnetic field. At high temperatures the ZFC and FC magnetizations match each other and gradually increase as temperature decreases, probably due to local clustering of the spins [15] or to ferromagnetic domain growth [14]. The ZFC and FC magnetizations start to differ below a certain field depended temperature $T_f(H)$, which is a phenomenon which can be interpreted in terms of spin-glass transition or freezing of domain walls motion [16]. The relation between $T_f$ and applied magnetic field is shown in the semilogarithmic plot in Fig. 2. It is evident that Almeida-Thouless (AT) line in the form of: $H=H_{AT}[1-T_f(H)/T_f(0)]^{3/2}$ where $H_{AT}$ is a coefficient, does not fit the experimental data correctly in whole explored field range i.e. from $3 \cdot 10^{-3}$ T to 3 T. The best fit to the AT line can be obtained for $\mu_0 H_{AT}=3.1$ T, $T_f(0)=71$ K and it matches the experimental data in the range of high magnetic fields, only. The data in the log($H$)-$T$ semilogaritmic representation, however, can be well fitted by a linear function, which means that $\log(H) \propto T_f(H)$. Therefore, the following relation is fulfilled between the value of applied magnetic field and $T_f$ temperature:

$$H = H_0 \exp[-bT_f(H)] \tag{1}$$

where $H_0$ and b are parameters of the fit equal to: $\mu_0 H_0$=4,7 T, b=-0.053 K$^{-1}$.

The relation described by eq. (1) can be explained within the model of domain-wall pinning proposed by Kersten [17] if one notices that the energy of a domain characterized by magnetic moment $m$ and interacting with a magnetic field is $E \propto mH$. Consequently eq. (1) can be expressed in terms of energy $E=E_0\exp[-bT_f(H)]$ which has identical form as the wall energy density $\sigma_w$=a·exp(-b$T$) [18]. In general, the pinning of domain-walls proceeds along a complex scenario in which the walls are bowed between adjacent pinning sites and can move to the next site when the applied field increases above a certain critical value. However, in the most simple approximation the domain-wall pinning energy $E_p$ is proportional to the wall energy density $E_p \propto \sigma_w$ [17] and the eq. (1) should be obeyed. Thus, in this approach the bifurcation between FC and ZFC magnetizations can be ascribed to the freezing of domain-wall motion below the field dependent temperature $T_f(H)$ and an increase in the strength of domain wall-pinning as the temperature further decreases. Above $T_f$ the energy due to thermal activation of the domain-walls exceeds the pinning energy and the walls can move between the pinning sites.

An effective test for glassy behaviour exhibited by many systems is the aging effect manifested as a magnetization dependence on time. The aging effect for our BFO sample is shown in Fig. 5, which presents the time dependence of thermoremanent magnetization after earlier cooling in the applied field $\mu_0 H$=0.1 T to a selected temperature followed by a rapid switching off this field. To better visualize the details of the magnetic moment relaxation vs. time, all curves $m(t)/m(t_0)$ in Fig. 5 are normalized with respect to the first recorded data $m(t_0)$ where $t_0 \approx 15$ s is the time necessary to switch off the magnetic field and to complete the first measurement.

When the data are plotted in $\log(m(t)/m(t_0))$ vs. $\log(t/t_0)$ form they co-ordinate linearly, except for the long term measurements for $T=80$ K. This means that the aging proceeds according to the following relation:

$$m(t) = m_\infty + m(t_0)\left(\frac{t}{t_0}\right)^{-A} \qquad (2)$$

where A is a coefficient.

The best fit to the experimental data in Fig. 5 within the model (2) is obtained for $m_\infty=0$ and A=0.033, 0.097 and 0.262 at 10, 40 and 80 K, respectively.

The power-law decay of the remnant magnetization $m(t)$ or aging as in eq. (2) can be an indication of FM domain development after rapid switch-off the field in the initially single domain FM state [19]. The decomposition of the homogeneous FM state into a polydomain one is probably controlled by the thermal fluctuations in a similar manner to the random-field controlled systems [20]. Indeed, identical dependences as eq. (2) have been theoretically obtained using Monte Carlo simulations in an assembly of single ferromagnetic domain nanoparticles with dipolar interparticle interactions [21] and found experimentally in superferromagnetic granular $Co_{80}Fe_{20}/Al_2O_3$ multilayers [19]. The negligible value of finite residual magnetic moment $m_\infty$ means that an intermediate FM domain concentration exists below some threshold above which the non-vanishing remanence is observed. A relaxation of $m(t)$ faster than predicted by the power-law, recorded at $T=80$ K for long-term measurements is probably caused by low pinning energy of domain walls, which at a temperature above $T_f(H)$ should be negligible (see Fig. 4)

Contrary to spin-glass state, the ferromagnetic or superferromagnetic systems controlled mainly by the processes of domain growth and domain walls pinning exhibit partial or negligible memory effect [14]. The result of test for memory effect for BFO is presented in Fig. 6. The measurements were performed in the following way: at first the reference curve (solid squares in Fig. 6) was determined on field cooling from room temperature down to 10 K at the cooling rate of 1 K·min$^{-1}$ and at the applied field $\mu_0 H$=0.1 T. Then, temperature was raised back and the procedure was repeated with cooling interrupted for $t$=10$^4$ s at $T_w$=50 K and the relaxation of the magnetic moment was measured (open squares in Fig. 6). After resuming the cooling and reaching 40 K, another measurement of magnetic moment $m$ was performed on heating the sample from 40 K to 60 K (open circles). The curve obtained on heating coincides with the reference line and does not exhibit any peculiarity at $T_w$. This result is in agreement with FM processes controlled by domain growth mechanisms which can erase the memory associated with domain wall pattern [14]. In contrast, for the spin-glass phase or domain wall reconformations (without domain growth) the memory effect should be present.

## 4. Conclusions

The results of our experiments, like the ZFC-FC bifurcation at the temperature of freezing of FM domain wall motion, power-law aging of magnetic moment and the absence of memory effect, seem to be well explained by the domain growth, domain wall pinning and reconformation processes in the BFO multiferroic. This possible explanation has been already mentioned in the report of Singh et al. [10] as a alternative to spin-glass, because many other systems can exhibit properties like AT-line occurrence (superparamagnets), ageing and rejuvenation (ferroic systems) or frequency dependent susceptibility (relaxors).

However, these two points of view do not exclude each other because the glassy behaviour can also appear due to reconformations of domain walls, which posses spin-glass like hierarchy of metastable states in real space [14]. Moreover, BFO is very weak FM with a rather low concentration of FM domains as evidenced by a negligible value of remnant magnetization $m_\infty$. Such systems are very close to transition between spin-glass and FM phase and can exhibit complex properties.

## Acknowledgements

This project has been granted by National Science Centre (project No. N N507 229040) and partially by COST Action MP0904.

References

[1]  G. Smolenskii, V. Yudin, E. Sher, and Y.E. Stolypin, *Antiferromagnetic properties of some perovskites*, Sov. Phys. JETP 16 (1963), pp. 622-624.

[2]  M. Fiebig, *Revival of magnetoelectric effect*, J. Phys. D: Appl. Phys. 38 (2005), pp. R123-R152.

[3]  A.M. Kadomtseva, Yu.F. Popov, A.P. Pyatakov, G.P. Vorobev, A.K. Zvezdin, and D. Viehland, *Phase transitions in multiferroic $BiFeO_3$ crystals, thin-layers, and ceramics: enduring potential for a single phase, room-temperature magnetoelectric "holy grail"*, Phase Transit. 79 (2006), pp. 1019-1042.

[4]  D. Lebeugle, D. Colson, A. Forget, M. Viret, P. Boville, J.F. Marucco, and S. Fusil, *Room-temperature coexistence of large electric polarization and magnetic order in $BiFeO_3$ single crystals*, Phys. Rev. B 76 (2007), pp. 024116-1-8.

[5]  T. Choi, S. Lee, Y.J. Choi, V, Kiryukhin, and S.-W. Cheong, *Switchable ferroelectric diode and photovoltaic effect in $BiFeO_3$*, Science 324 (2009), pp. 63-66.

[6]  D. Khomskii, *Classifying multiferroics: Mechanisms and effects*, Physics 2 (2009), pp. 20- 28.

[7]  A. Lubk, S. Gemming, and N.A. Spalin, *First-principles study of ferromagnetic domain walls in multiferroic bismuth ferrite*, Phys. Rev. B 80 (2009), pp. 104110-1-9.

[8]  C. Ederer and N.A. Spaldin, *Influence of strain and oxygen vacancies on the magnetoelectric properties of multiferroic bismuth ferrite*, Phys. Rev. B 71 (2005), pp. 224103-1-11.

[9]  M.K. Singh, W. Prellier, M.P. Singh, R.S. Katiyar, and J.F. Scott, *Spin-glass transition in single-crystal $BiFeO_3$*, Phys. Rev. B 77 (2008), pp. 144403-1-5.

[10] M.K. Singh, R.S. Katiyar, W. Prellier, and J.F. Scott, *The Almeida-Thouless line in $BiFeO_3$: is bismuth ferrite a mean field spin glass?*, J. Phys.: Condens. Matter. 21 (2009), pp. 042202-1-5.

[11]   M.K. Singh, S. Dussan, W. Prellier, and R.S. Katiyar, *One-magnon light scattering and spin-reorientation transition in epitaxial BiFeO$_3$ thin films*, J. Magn. Magn. Matter. 321 (2009), pp. 1706-1709.

[12]   I. Sosnowska, T. Peterlin-Neumaier, and E. Steichele, *Spiral magnetic-ordering in bismuth ferrite*, J. Phys. C: Solid State Phys. 15 (1982), pp. 4835-4846.

[13]   M. Ramazanoglu, M. Laver, W. Ratcliff II, S.M. Watson, W.C. Chen, A. Jackson, K. Kothapalli, S. Lee, S.-W. Cheong, and V. Kiryukhin, *Local weak ferromagnetism in single-crystalline ferroelectric BiFeO$_3$*, Phys. Rev. Lett. 107 (2011), pp. 207206-1-5.

[14]   E. Vincent, V. Dupuis, M. Alba, J. Hammann, and J.-P. Bouchaud, *Aging phenomena in spin-glass and ferromagnetic phases: Domain growth and wall dynamics*, Europhys. Lett. 50 (2000), pp. 674-680.

[15]   S. Nakamura, S. Soeya, N. Ikeda, and M. Tanaka, *Spin-glass behavior in amorphous BiFeO$_3$*, J. Appl. Phys. 74 (1993), pp. 5652-5657.

[16]   H. Chang, Y.-q. Guo, J.-k. Liang, and G.-h. Rao, *Magnetic ordering and irreversible magnetization between ZFC and FC states in RCo$_5$Ga$_7$ compounds*, J. Magn. Magn. Matter. 278 (2004), pp. 306-310.

[17]   M. Kersten, *Über die Bedeutung der Versetzungsdichte für die Theorie der Koerzitivkraft rekristallisierter. Werkstoffe*, Z. Angew. Phys. 8 (1956), pp. 496-502.

[18]   G. Vertesy, and I. Tomas, *Temperature dependence of the exchange parameter and domain-wall properties*, J. Appl. Phys. 93 (2003), pp. 4040-4044.

[19]   X. Chen, W. Kleemann, O. Petracic, O. Sichelschmidt, S. Cardoso, and P.P. Freitas, *Relaxation and aging of a superferromagnetic domain state*, Phys. Rev. B. 68 (2003), pp. 054433-1-5.

[20]   Y. Imry, and S.K. Ma, *Random-field instability of ordered state of continuous symmetry*, Phys. Rev. Lett. 35 (1975), pp. 1399-1401.

[21]   M. Urlich, J.G.-Otero, J. Rivas, and A. Bunde, *Slow relaxation in ferromagnetic nanoparticles: indication of spin-glass behavior*, Phys. Rev. B. 67 (2003), pp. 024416-1-4.

Figure Captions

Figure 1 XRD-pattern the BiFeO$_3$ bismuth ferrite micro-cubes. The asterisk * indicates traces of Bi$_{25}$FeO$_{40}$ parasitic phase.

Figure 2 SEM micrograph of the BiFeO$_3$ bismuth ferrite micro-cubes synthesized by means of microwave activation of the Pechini process.

Figure 3. ZFC and FC magnetization temperature dependences recorded for three different magnetic fields: 0.01 T; 0.1 T and 1 T. The arrows indicate the temperature $T_f$ at which the difference between ZFC and FC magnetization appears.

Figure 4. Relation between the temperature $T_f$ at which the difference between FC and ZFC magnetizations appears vs. the applied magnetic field. Solid red line corresponds to the domain walls pinning model equation (1), the broken blue line represents the Almeida-Thouless transition.

Figure 5. Normalized value of magnetic moment relaxation vs. normalized time (aging) after field cooling the sample in the applied magnetic field $\mu_0H$=0.1 T to 10 K; 40 K and 80 K

Figure 6. The absence of memory effect in BFO. Solid squares represent the reference magnetization measured on field-cooling in $\mu_0H$=0.1 T. Open squares represent FC magnetization measurement on cooling with intermittent stop at 50 K for $t_w$=10$^4$ s. Open circles correspond to FC magnetization measured when warming from 40 K to 60 K.

Figures

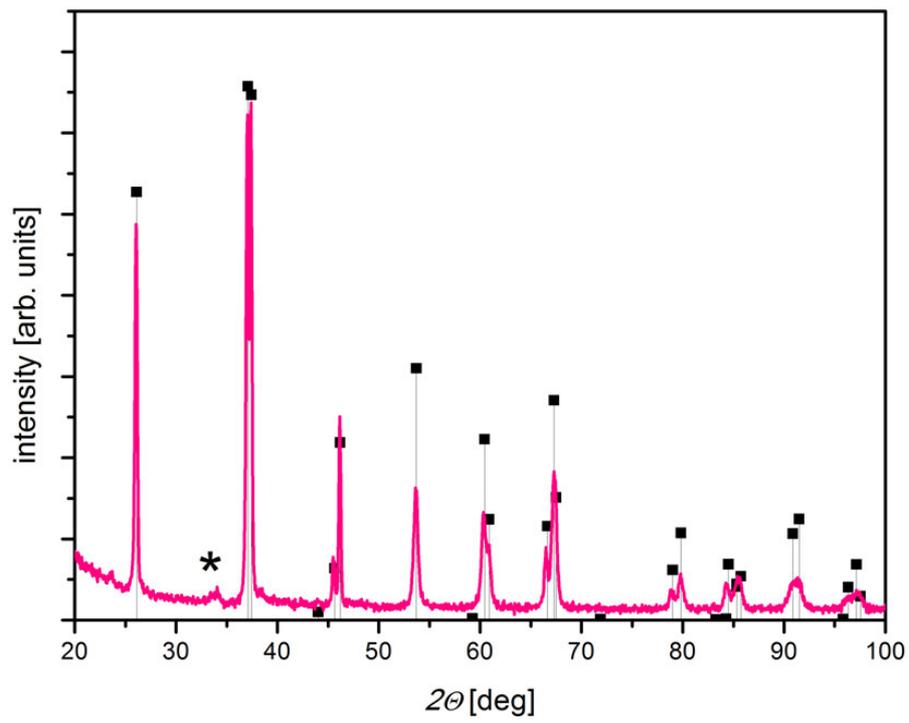

Figure 1

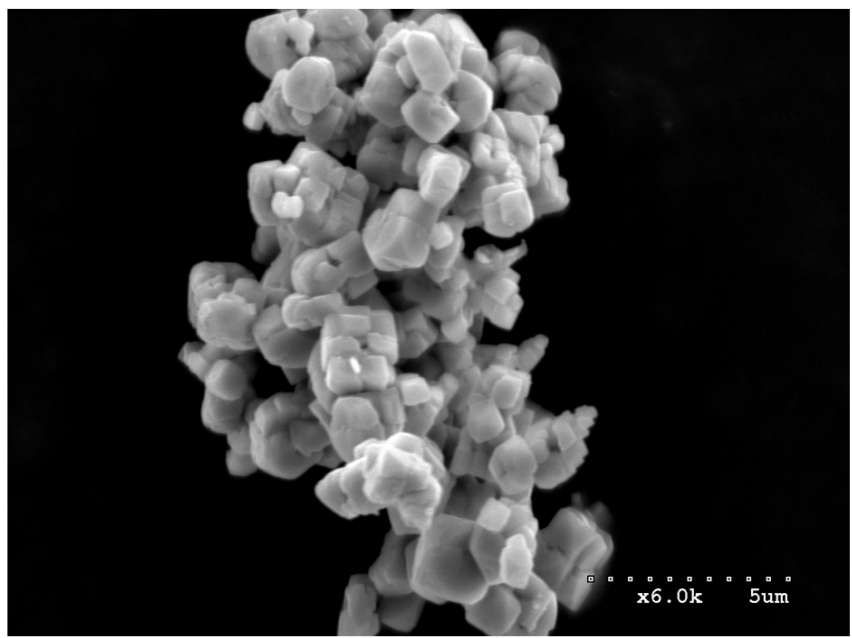

Figure 2

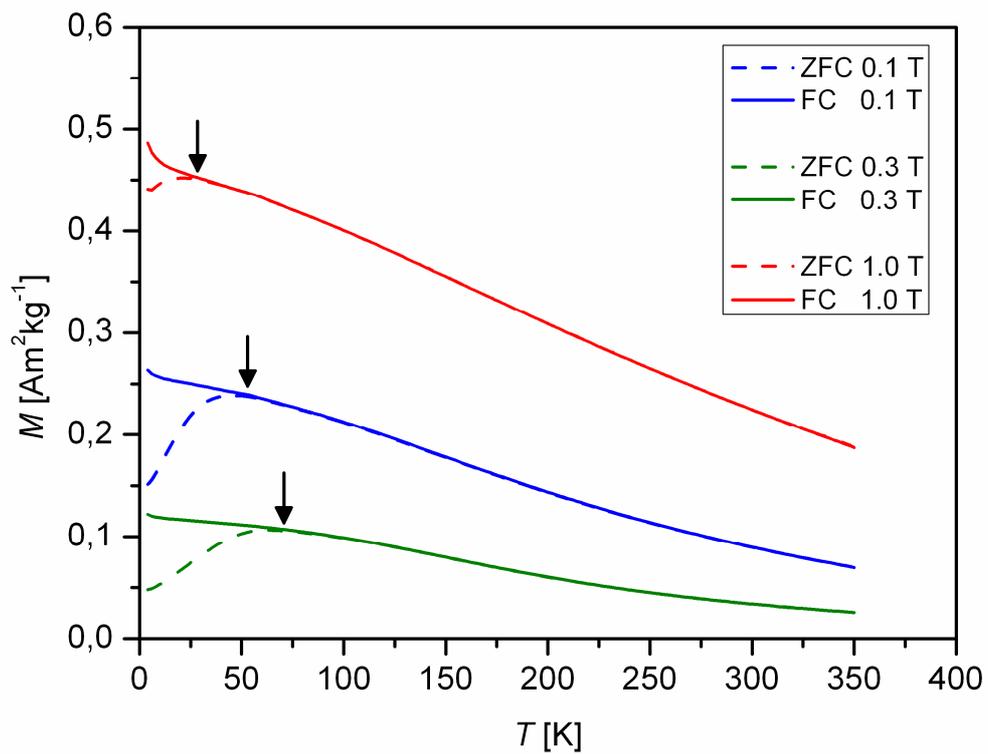

Figure 3

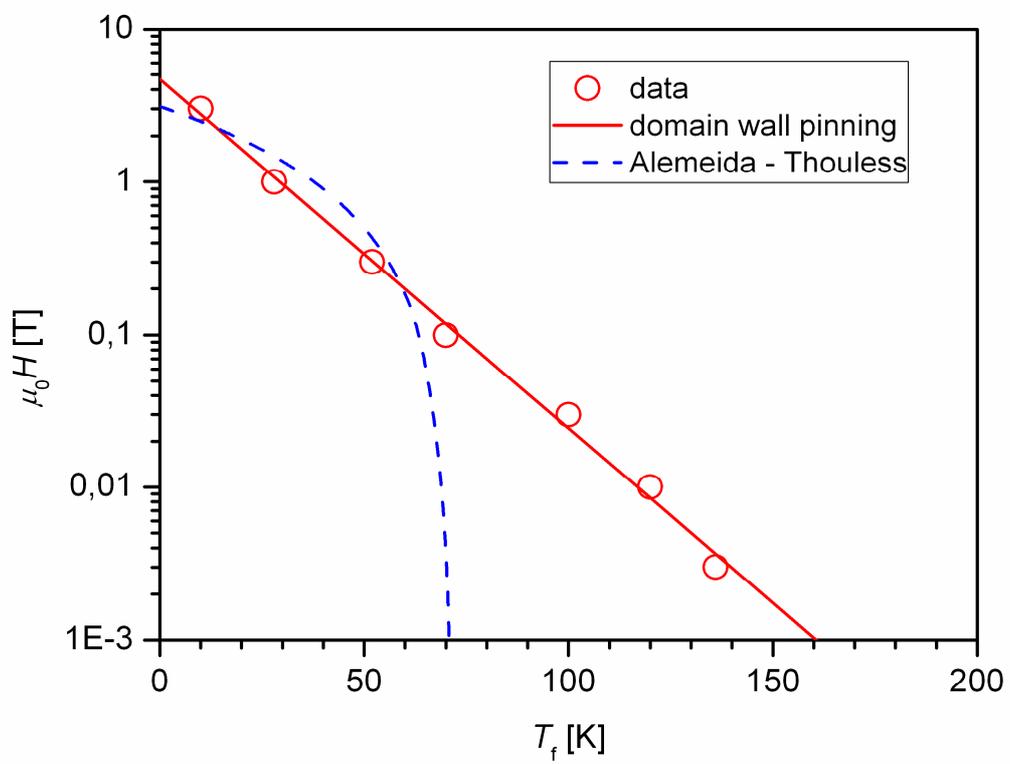

Figure 4

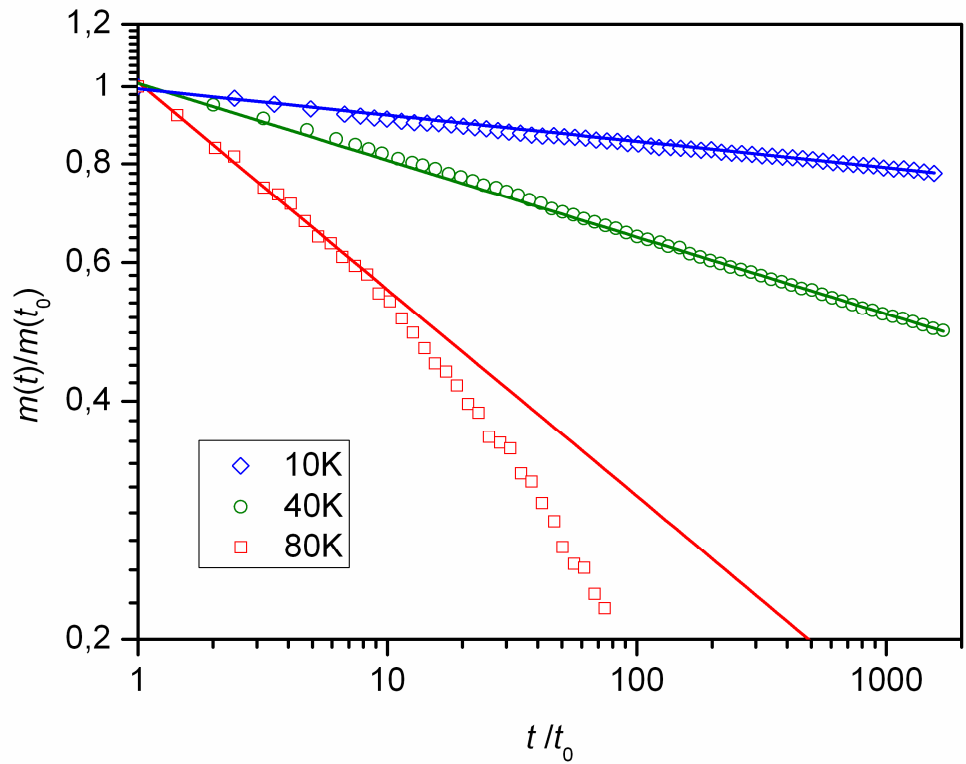

Figure 5

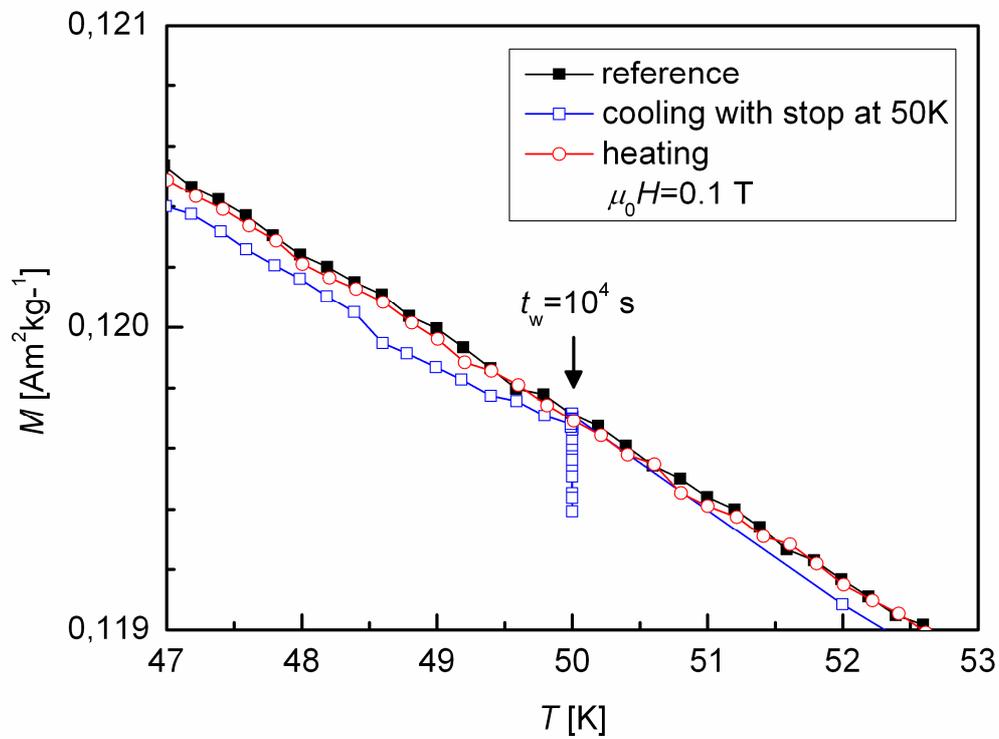

Figure 6